\documentclass[aps,prb,amssymb,amsmath,showpacs,superscriptaddress]{revtex4-1}
\usepackage{graphicx}
\usepackage{longtable}

\begin{document}
\title{Experimental Quantification of Entanglement Through Heat Capacity}

\author{H. Singh}

\author{T. Chakraborty}

\author{D. Das}

\affiliation{Indian Institute of Science Education and Research (IISER) Kolkata,
Mohanpur Campus, PO: BCKV Campus Main Office, Mohanpur – 741252, Nadia, West Bengal, India.
}
\author{H.S. Jeevan}
\author{Y. Tokiwa}
\author{P. Gegenwart}
\affiliation{I. Physik. Institut, Georg-August Universit\"{a}t, G\"ottingen, D-37077 G\"ottingen, Germany.}
\author{C. Mitra{*}}
\affiliation{Indian Institute of Science Education and Research (IISER) Kolkata,
Mohanpur Campus, PO: BCKV Campus Main Office, Mohanpur – 741252, Nadia, West Bengal, India.
}

\email{chiranjib@iiserkol.ac.in}


\begin{abstract}
A new experimental realization of heat capacity as an entanglement witness (EW) is reported. Entanglement properties of a low dimensional quantum spin system are investigated by heat capacity measurements performed down to very low temperatures (400mK), for various applied magnetic field values. The experimentally extracted results for the value of heat capacity at zero field matches perfectly with the theoretical estimates of entanglement from model Hamiltonians. The studied sample is a spin $\frac{1}{2}$ antiferromagnetic system which shows clear signature of quantum phase transition (QPT) at very low temperatures when the heat capacity is varied as a function of fields at a fixed temperature. The variation of entanglement as a function of field is then explored in the vicinity of the quantum phase transition to capture the sudden loss of entanglement.
\end{abstract}
\maketitle

\section{Introduction}
Quantum communication between two subsystems, require establishment of genuine quantum correlation between them\cite{NielsenChuang, EPR}. These quantum correlations manifest through entanglement, 
which is the most fundamental physical resource for quantum information processing \cite{vedral0}. 
Hence, it is important to quantify experimentally the amount of entanglement present in the system under consideration using entanglement witness (EW) operators. Archetypal EW operators are physical observables, which can not only distinguish between separable and entangled states, but also quantify the entanglement content \cite{vedral2, vedral1, wootters1}.

Correlation energy of any composite system, consisting of identical particles, either bosonic or fermionic, due respectively to the symmetric or anti-symmetric nature of the ground state wave functions, is well known to affect the ground state energy \cite{griffith}. Symmetric or anti-symmetric wave functions do not generally allow separability, yielding naturally occurring entangled state. Internal energy (U), which in turn is related to observables, such as susceptibility and heat capacity, carries information about the ground state of the system. Not only can they measure the ground state, but can also reveal when it transits from one type of symmetry to the other, changing the value of `U'. This is especially the case where one sweeps a parameter that varies adiabatically. For example, in correlated electrons, often the ground state is a spin singlet state, yielding a particular value of `U'. When one varies the magnetic field, there can be a crossover to a triplet like state, which has a different symmetry accompanied by a change in `U'. Thus, `U' can be used as a marker in classifying the nature of the ground state. It can probe the ground state symmetry, thereby distinguishing between separable and entangled states. 

 
Entanglement study in many body systems, like spin chains \cite{bose01} have received tremendous interest because of the presence of naturally occurring stable entangled states \cite{vedral0, bose2}. Quantification of entanglement in spin systems have been reported by several groups \cite{indrani, sarthour1, rappoport, sarthour2, dipta}. Also, entanglement study in such systems will be of fundamental importance for following reasons, (i) to determine limits of entanglement in terms of size, i.e., how large macroscopic system can exhibit entanglement, (ii) to check the robustness of entanglement with temperature,  (iii) to explore whether entanglement can be an order parameter to capture quantum phase transition, (iv) Designing new materials for quantum information processing, as entanglement is an essential resource for most quantum protocols \cite{vedral0}. Thermodynamic observables like magnetic susceptibility \cite{sarthour1, sarthour2, dipta} and neutron scattering \cite{bruckner} have shown potential importance for entanglement estimation in macroscopic systems.

A new experimental study, making use of heat capacity as EW to quantify the entanglement content is reported. For this purpose, we have chosen copper nitrate (CN) $Cu(NO_{3})_{2}\cdot2.5 H_{2}O$ single crystals, which is regarded as the first inorganic spin chain compound investigated experimentally \cite{bonner}.This compound 
is captured by the Hamiltonian
\begin{equation}
\label{hamiltonian}
H = J^{\prime} \sum_{i} ( S_{i} \cdot S_{i+1} + \alpha (S_{i+1} \cdot S_{i+2}))+ g\mu_B B \sum_{i} S_{i}^z,                                        
\end{equation}
where, alternation parameter $\alpha = 0.27$ \cite{bonner}. 
Also, $J^\prime = 4J$ and J is the exchange coupling constant. 
Since, the inter-dimer coupling is weak, the ground state turns out to be set of dimers which are very weakly coupled or uncoupled dimers \cite{xu, vedral2}. In our measured temperature regime, CN exhibits dimerization with very weak inter-dimer coupling, which on the other hand becomes important only at very low temperature ($<100mK$) \cite{bonner}.
For the antiferromagnetic case, the observable corresponding to the staggered magnetization, $ M_s  = \dfrac{1}{N}(\sum_{i} (-1)^i S_{i}^{z})$, does not commute with the Hamiltonian in the absence of external magnetic field. Thus,  $M_s$ exhibits quantum fluctuations. This non-commutativity is fundamentally responsible for the ground state to be entangled and the corresponding entangled ground state for the dimerized system is well represented by the nearly maximally entangled singlet state ($\mid\psi^{-}\rangle = \frac{1}{\sqrt{2}} [\mid\uparrow\rangle_{1} \mid\downarrow\rangle_{2} - \mid\downarrow\rangle_{1} \mid\uparrow\rangle_{2}]$) with an associated energy of $-3J^\prime/4$ \cite{mitra1}. In this work, all the analysis was done successfully by considering a four dimensional Hilbert space. Theoretical study of entanglement in this system has been reported by Brukner \textit{et al.} \cite{bruckner}, where they considered the data of \cite{xu, berger}. Brukner \textit{et al.} have successfully demonstrated validity of dimer model in estimation of entanglement in CN using experimental neutron scattering and magnetic susceptibility as an EW \cite{bruckner, vedral2}. They considered the neutron scattering data of Xu \textit{et al.} \cite{xu}, which itself shows good match with the theoretical dimer model.
It was reported that macroscopic observables like susceptibility \cite{bruckner,vedral2} and internal energy \cite{bandyo, vedral3, dowling, toth, wang} can detect entanglement between spins \cite{bose2, vedral2}. 
Das \textit{et al.} have quantified entanglement from experimental magnetic susceptibility of CN and have shown its variation with field and temperature \cite{dipta}.
Susceptibility as an EW, however, has certain limitations because of its applicability to magnetic systems only \cite{vedral2}. However, on the other hand internal energy as EW is believed to encompasses wide range of systems. T\`{o}th \textit{et al.} \cite{toth} and Wang \textit{et al.} \cite{wang}, have discussed in detail the estimation of entanglement from internal energy over wide range of model systems.

Recently, it was shown by Wie\'{s}niak \textit{et al.} \cite{vedral1} that heat capacity can also detect entanglement in a system.  Measurement of heat capacity is a standard and well established experimental protocol. Here, we report  the  experimental realization of heat capacity as a new entanglement witness (EW). We quantitatively extracted  entanglement from specific heat measurements
. These measurements were carried out in a field regime,
well above the critical field to ferromagnetically polarize the system. This allows us to exhaustively investigate the quantum phase transition (QPT) \cite{subir, philipp}, a phase transition driven by quantum fluctuations
. At the quantum critical point (QCP) the behaviour of entanglement has been theoretically studied  extensively by several groups \cite{osborne, osterloh, vidal, venuti,song}. They have found that the entanglement displays scaling behaviour close to the QCP. Hence, from the perspective of entanglement it is worth studying its behaviour experimentally in the vicinity of QCP. We are able to directly probe the quantum phase transition at the field of 3.5T in this system. Grenier \textit{et al.} has thoroughly investigated QPT in this system through inelastic neutron scattering and has unveil that QPT in this system has a characteristics of Bose-Einstein condensation of magnons \cite{grenier}. A more detailed discussion on the nature of QPT in this system can be found in an earlier work \cite{dipta}.   \\

\section{Experimental Details}
The heat capacity measurement was done on the 99.999\% pure CN sample procured from Sigma Aldrich, in the Quantum Design physical properties measurement system (PPMS) 
. The heat capacity vs. temperature measurements, in the range 400mK to 15K, was performed at various applied magnetic field values, varying from 0T to 7T. A carefully measured amount of sample was taken and the data was subsequently corrected by the addenda measurement performed before loading the sample.

\section{Results}

Wie\'{s}niak \textit{et al.} \cite{vedral1} considered the extraction of entanglement from theoretically generated heat capacity curves for certain Hamiltonians, whose variance can be minimised over separable states, as has been done for the transverse field Ising model. In the present case, for the Heisenberg Hamiltonian, this approach does not work, since a separable state can also be an eigenstate of Heisenberg Hamiltonian \cite{dipta}. For a thermal state, we explicitly know the energy eigenvalues and value of the variance $\Delta^2 H$,
which is $(\langle H^2 \rangle - \langle H \rangle^2) $. 
So, we have made use of a different procedure to handle this complexity.  We have quantified entanglement through internal energy, which could be used as an EW in addition to magnetic susceptibility, especially for low dimensional spin systems. The internal energy, when expressed in terms of specific heat, is given as, $U = U_{\circ} + \int C_{p}(T) dT$, where $U_{\circ} $ being the ground state energy. For a dimerized spin system, the Concurrence \textit{C}, the measure of entanglement for a mixed state, is given by \cite{wootters1, wootters2},

\begin{equation}
C = \frac{1}{2} max [0, \frac{\vert U \vert}{NJ} - 1].
\label{EW1}
\end{equation}
Here, N is the number of qubits. The concurrence for a bipartite system is a good measure of entanglement and has successfully led to an estimation of entanglement for this particular system \cite{dipta}. 

\begin{figure}[b]
\includegraphics[scale=0.08]{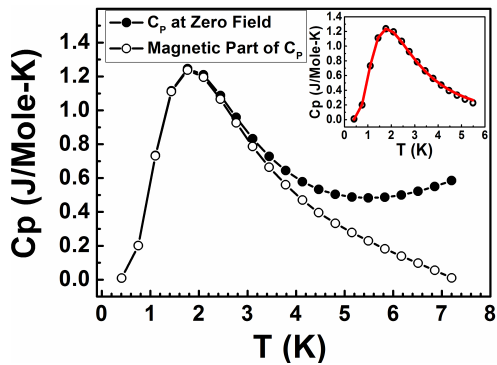}
\centering
\caption{The unprocessed data of heat capacity vs temperature for CN in the absence of field shown by (solid 
circles) and the magnetic contribution 
by (open 
circles)
. The inset shows fit of magnetic part
to the dimer model Hamiltonian.}
\label{raw_data}
\end{figure}

The heat capacity of CN measured in the absence of field is shown in Fig. \ref{raw_data}. The magnetic part of heat capacity, used for entanglement analysis, was extracted by a careful subtraction of the lattice part ($\beta' = 1.1 \times 10^{-3} J/Mole-K^{4}$) \cite{friedberg}. 
In the inset of Fig. \ref{raw_data}, the magnetic part of zero field heat capacity data is plotted along with the theoretical curve generated using the dimer model Hamiltonian. As we can see, the two curves fall on each other, suggesting a good match. The exchange coupling $\dfrac{J^\prime}{k_B} $ has been considered to be around 4K \cite{dipta}. 
In the heat capacity vs. temperature curve (Fig.\ref{raw_data}), an anomaly is observed around 2K, known as Schottky anomaly \cite{esr}, which is  typical of a gapped two-level system. In the zero field scenario, at very low temperatures, the system is in its ground state and has very low probability of transition to the excited state. As a result there is almost no change in internal energy with temperature, resulting in low heat capacity because heat capacity is the differential change in internal energy with temperature. As the temperature is raised and is equal to energy gap between the ground state and the first excited state, it starts absorbing energy and a broad maximum in heat capacity is observed as the temperature in varied. The broad maximum in this case occurs around 2K. At higher temperatures all the relevant states are equally populated giving no differential change in internal energy with temperature, hence the heat capacity drops. 

\begin{figure}[t]
\centering
\includegraphics[scale=0.25]{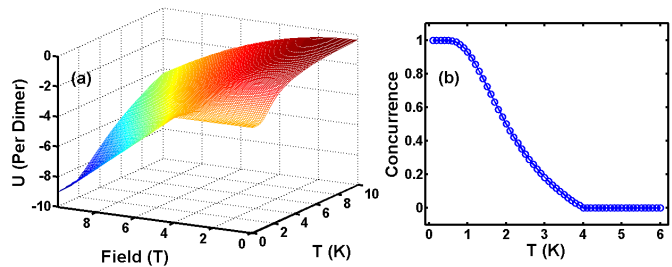}
\caption{(a)A surface plot exhibiting `U', temperature and magnetic field along three axes 
. (b)Theoretically extracted entanglement at zero field.} \label{theory_U}
\end{figure}

In order to match with the EW, we determined `U' of the system, which is the expectation value of the Hamiltonian, $U = \langle H \rangle$. The 3D plot of theoretically generated curves of `U' are shown in Fig. \ref{theory_U}(a). These curves were generated numerically, where we considered the thermal density matrix
$\rho_{T} =  \frac{1}{Z}  e^{-\beta H}$ \cite{dipta}. Here, $\beta = \frac{1}{k_{B} T} $
and $Z = Tr(\rho) = e^{ \frac{3J^\prime}{4}\beta}+  e^{ \frac{-J^\prime}{4}\beta} + e^{(\frac{-J^\prime}{4}+ g \mu_{B} B) \beta} + e^{- ( \frac{J^\prime}{4}- g \mu_{B} B)\beta}$, with $U = Tr( \rho_{T} H)$.
The detailed analysis can be found in Das \textit{et al.} \cite{dipta}. The value of U, shown in the 3D plot [Fig. \ref{theory_U}(a)], is normalized to the value of one dimer. One can see that `U' in zero fields saturates to a value of -3 at zero temperature. It is apparent from the zero temperature curves that, with the increase in field, `U' remains constant up to a certain critical value, beyond which there is a sharp increase, which reaches up to a value of -9.
This is an indicator of a quantum phase transition, that occurs due to a level crossing of two of the lowest energy eigenstates, beyond a critical field. It is worth mentioning here that, heat capacity is a bulk measurement and hence is representative of the thermodynamic limit, even though the spins dimerize, presence of very weak inter-dimer coupling yields a many body ground state and hence QPT is observed. Below the critical field (3.5T), 
the system is in an antiferromagnetic singlet state, whereas above the critical field, the resulting ground state is the separable triplet state $(\mid\upuparrows\rangle)$ with the associated energy $ \dfrac{J^\prime}{4} - g \mu_{B}B$. The effect is sharp at lower temperatures, in which the system is predominantly in the ground state.
On increasing the temperature the system is no longer in a pure state but in a mixture of states, as a consequence of which the change in `U' as a function of field is not abrupt but more gradual. 
Using Eq. 2, the concurrence at zero field has been extracted and is shown in Fig. \ref{theory_U}(b). It shows a decrease in entanglement with temperature which disappears at 4K.

\begin{figure}[t]
\includegraphics[scale=0.1]{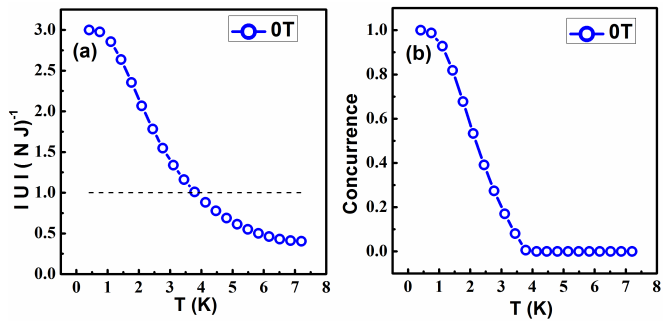}
\centering
\caption{(a) Plot of internal energy vs temperature with the bound as mentioned in the text.
(b) Extraction of entanglement from experimental curve in the absence of field}
\label{Exp_U}
\end{figure}

Experimentally, `U' at zero field is calculated by carrying out numerical integration on zero field heat capacity data, incorporating theoretically evaluated $U_{\circ}$. 
The condition for separability at zero field gives a bound for `U',
\begin{equation}
\frac{\vert U \vert}{NJ} > 1
\end{equation}
such that the states with $ \frac{\vert U \vert}{NJ} < 1 $ are separable states and the ones above 1 (dashed line in Fig. \ref{Exp_U}(a)) are entangled. As mentioned above, concurrence is a good measure of entanglement for a bipartite system and hence can be well implemented for non-interacting dimer system using Eq. \ref{EW1}. In our measured temperature regime, each dimer can be considered to have very weak interaction with neighbouring dimers. 
The experimental results clearly show entanglement in the system. 
The experimentally estimated values of entanglement here (Fig.\ref{Exp_U}(b)) are comparable to the theoretical values of entanglement (Fig.\ref{theory_U}(b)) generated using the measure (concurrence), postulated by Wootters \textit{et al.} \cite{wootters1}.
We see a similar behaviour, corroborating specific heat as a reliable EW. The region of the curve above the dashed line [Fig. \ref{Exp_U}(a)] is not separable and hence can not be explained without entanglement.
The extraction of entanglement for this curve at zero fields also matches reasonably well with the theoretical prediction, as shown in Fig. \ref{Exp_U}(b). At low temperatures, the system is predominantly in its nearly maximally entangled ground state, represented by $\vert\psi^{-}\rangle$. One can see that with increase in temperature the entanglement decreases. This is because with increase in temperature the contribution from all the three triplet state increases at the expense of the singlet ground state. At higher temperatures, the system is approximately an equal mixture of all the states, where it can be shown that the entanglement is completely lost \cite{mitra1, dipta}. 

\begin{figure}[b]
\includegraphics[scale= 0.8]{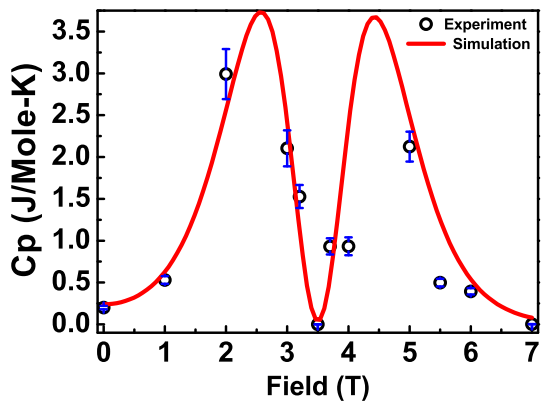}
\centering
\caption{Variation of experimental heat capacity with field at 800mK and theoretically simulated curve for the dimer model 
plotted on top of each other. The corresponding error bars were also shown.}
\label{QPT}
\end{figure}

 We now explore the variation of heat capacity as well as entanglement extracted out of this data as a function of field. For this we extracted the values of heat capacity at 0.8K. These extracted values plotted as a function of field are shown in Fig. \ref{QPT}. 
The minima at 3.5T is a Schottky like anomaly, but as a function of field. At very low fields, the system is in the ground state ($ \vert\psi^{-}\rangle $). 
The population of this ground state is $ e^{\frac{-3J^\prime\beta}{4}}$ and that of the first excited state is  $ e^{-(\frac{J^\prime}{4}- g \mu_{B} B)\beta} $. As the field increases, the first excited state $ (\mid\upuparrows\rangle) $ approaches the ground state and the system starts absorbing thermal energy from the environment  as the gap $ \Delta = E_{e1} {-} E_{g} $ decreases; thus the heat capacity increases. However, as the first excited state approaches the ground state, its population increases at the cost of reduction in population of the ground state. This reduces the absorption as the number of electrons now available for excitation from the ground state of the ensemble goes down gradually. Thus the heat capacity starts dropping after reaching a maximum. At the QCP, the two states are equally populated since they have the same energy. 
Hence, there is no excitation as the next energy level ($ \vert\psi^{+}\rangle = \dfrac{1}{\sqrt{2}} (\vert\uparrow\downarrow\rangle + \vert\downarrow\uparrow\rangle) $) is at $(\dfrac{J^\prime}{4} - g \mu_{B} B)$; the system does not absorb any more energy. The heat capacity tends to zero, at the level crossing. Upon further increasing the field, \textit{i.e.}, beyond the level crossing point, the heat capacity starts to rise again, as seen in our data.  This is because, when  we increase the field beyond the QCP, the gap between the current ground state and the erstwhile ground state (which is now the current first excited state), starts increasing again. In addition to that, with increasing field, the population of the new ground state $ (\mid\upuparrows\rangle) $ goes on increasing. Following the similar argument, the heat capacity reaches a second maxima and then drops to zero when the gap between the new ground state and the new first excited state $ ( \vert\psi^{-}\rangle ) $ is larger than the excitation energy available from the environment. 
In our case, interestingly, this dip which is present at very low temperature dies out with increase in temperature, confirming this truly is a quantum phase transition \cite{dipta}. 

We have simulated the variation of heat capacity with field  using the relation,
\begin{equation}
C_{p} = \frac{1}{k_{B}T^2} (\langle H^2 \rangle - \langle H \rangle^2 )
\label{Cp_theory}
\end{equation}
(assuming $J^\prime$ = 4) where, H is the Heisenberg Hamiltonian. This variation is shown in Fig. \ref{QPT}. One can see that the simulated curve and the experimental results are in good agreement with each other. Like susceptibility, heat capacity is also a response function and hence measures the non-local properties of the system, which makes this measurement suitable for studying entanglement, which occurs as a result of non-local interaction between spins of the system. 

We must mention that since we are at finite temperatures and not at absolute zero, there will be some contribution of thermal fluctuation to the QCP. The QCP can be approached in 2 ways, either by approaching the QCP as a function of temperature by sitting close to the quantum critical point (in parameter space) or by sitting at a temperature $T << J$, and varying the magnetic field \cite{subir}. However, in our experiments we have measured down to 0.4K, which is quite low in comparison to “J=4K”, hence the QCP is not significantly affected by thermal fluctuation and what we are observing is indeed QPT. Being in the temperature range 0.4K to 0.8K, rules out the effect of thermal fluctuations to a large extent. However, if one cools the system  down to 10mK, one could completely get rid of any thermal fluctuation whatsoever and will be able to observe a very sharp transition at QCP, as can be seen from the lowest temperature curve of the theoretical 3D plot of Fig.7(a) of Ref.\cite{dipta}. In this earlier work \cite{dipta}, we could successfully capture the QPT as a variation of magnetic field up to 7T from purely magnetic measurements.
\begin{figure}[t]
\centering
\includegraphics[scale=0.9]{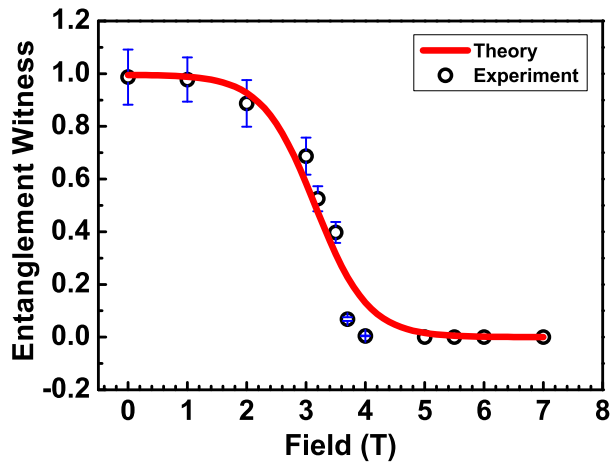}
\caption{Plot of theoretically simulated (solid curve) and experimentally quantified (open circles) depicting variation of entanglement with field at 0.8K. The error bar is also shown along with the data. 
}
\label{ESL}
\end{figure}
It was achieved using the two complementary observables Q and P, where P describes the local properties of individual spins, Q the non-local quantum correlations between spins \cite{vedral2}. For this, one has to measure both the susceptibility (which is used to generate Q) as well as magnetization (which is used to generate P) at each temperature and field values.  However, in the present case we have captured QPT by directly measuring only one observable, the heat capacity. This interesting result has motivated us to explore entanglement behaviour in the vicinity of the QCP. We have used the theoretically generated plots of internal energy for the Heisenberg Hamiltonian (assuming $J^\prime$ = 4) to generate the internal energy as a function of field. We then use the EW
\begin{equation}
C = \frac{1}{2} max [0,\frac{\vert U-BM \vert}{NJ}- 1],
\label{EW2}
\end{equation}

\begin{figure}[t]
\centering
\includegraphics[scale=0.08]{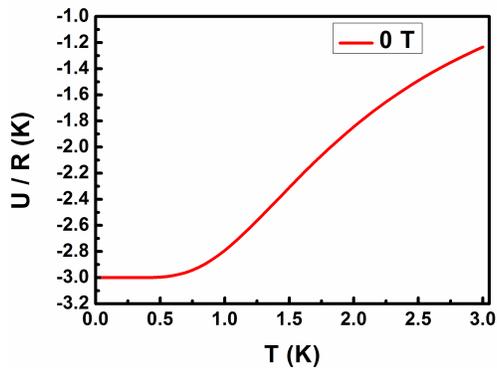}
\caption{Theoretical variation of internal energy with temperature for zero applied magnetic field.} 
\label{theory internal}
\end{figure}
to capture the variation of entanglement with magnetic field at 0.8K. Here, $ M = Tr[\rho (S_{1}^z + S_{2}^z)] $ is the magnetization. This behaviour is depicted in Fig. \ref{ESL}. The sharp drop indicates the QCP. The variation of entanglement as a function of temperature and field is very similar to the one generated by Arnesen \textit{et al.} [Fig 1, Ref. \cite{bose2}]. To verify this result experimentally, we used our experimental heat capacity data to calculate the internal energy as a function of field. This was done using numerical integration, where $U_{\circ}$ was calculated theoretically for every field values at a fixed temperature (which in the present case is 0.8K). The value of $U_{\circ}$ was calculated theoretically using Heisenberg Hamiltonian for a dimer model (assuming $J'$=4). We generated the variation of internal energy with temperature for a fixed magnetic field value. It was found that in the low temperature regime, for zero field curve (Fig. \ref{theory internal}), the variation of internal energy with temperature is almost constant and there is hardly any difference between internal energy values at 0.8K and ground state energy ($\dfrac{-3J'}{4}$). The eigenstate corresponding to this ground state ($-3J'/4$), is the maximally entangled singlet state $\vert\psi^{-}\rangle$. However, at different applied magnetic fields the variation of internal energy with temperature is different \citep{dipta}. Since 0.4K is significantly less than $J'$(4K), $U_{0.4}$ is $-3J'/4$ for zero field data. Hence, we incorporated different $U_{\circ}$ at fixed temperature (0.8K), to calculate internal energy for different applied magnetic field values. Then we used the same witness operator Eq. \ref{EW2} to calculate the entanglement with field at 0.8K as shown in Fig. \ref{ESL}. We observed a striking match between theoretical prediction and experimental results displaying a sudden drop in entanglement when field is increased beyond the critical value.
Though, in the present work we have studied a dimer/alternating dimer system, heat capacity as an EW can be extended to spin chains as well as ladders. Heat capacity being a connected two point correlation function can measure non-local correlation not only amongst neighbouring spin but also between spins that are separated by the relevant correlation length ($\xi$). For a many body systems,  $C_P = \frac{1}{k_B T^2} \sum_{i,j} (\langle S_i \cdot S_j \rangle - \langle S_i \rangle \langle S_j \rangle)$, where, $\vert i-j \vert \leq \xi$. In such a chain or ladder like systems, the entanglement measured would be an averaged out entanglement emanating from all entangled pairs which need not be only the nearest neighbour pairs. The nearest neighbouring pairs would of-course exhibit maximum entanglement.

Many 1D spin system, are known to be connected with conformal field theory. The conformal charge characterizing these systems is connected with specific heat as well as measure of entanglement through fluctuations \cite{blote,song}. Heat capacity being a measure of two point correlation function can capture these fluctuations. This provides an experimental access to non-local correlation in the form of entanglement. The connection of entanglement entropy with fluctuation is well established for simple systems like free fermions \cite{song} . The study of more complicated models particularly from experimental point of view is hence desirable. Also, it is worth exploring in future from theoretical as well as experimental point of view, the relation between fluctuation and entanglement entropy for gapped spin half systems. Since the correlation length of gapped spin systems is finite as opposed to gapless systems where, correlation lengths could be very large, it would be interesting to explore how the fluctuations affect the entanglement entropy of a gapped spin half system.

\section{Conclusion}
In conclusion, we have studied the heat capacity of CN sample, a spin $\frac{1}{2}$ dimer system, possessing an entangled singlet ground state, down to 400mK. We have successfully quantified the entanglement from experimental heat capacity data,  demonstrating that heat capacity can be a reliable EW. The entanglement extraction from the heat capacity data has shown reasonably good match with the theoretically generated values. We have studied the variation of entanglement with temperature, where the entanglement reduces with rise in temperature. 
We have also studied the variation of heat capacity with magnetic field and were able to observe quantum critical behaviour in the vicinity of 0.8K which is a reasonably low temperature. The observation of a Schottky like anomaly (Fig. \ref{QPT}) conclusively demonstrates the presence of a QCP due to level crossing. Finally, we observed a sharp drop in entanglement witness as one varies the applied magnetic field beyond the critical field.

\section{Acknowledgement}
The authors would like to thank the Ministry of Human Resource and Development, Government of India and G\"{o}ttingen Kolkata Open Shell Systems (G-KOSS) for funding. We also thanks Prof. P.K. Panigrahi and Prof. Andreas Honecker for useful discussions and suggestions.


\begin{thebibliography}{99}

\bibitem{NielsenChuang} Nielsen M A and Chuang I L 2000 \textit{Quantum Computation and Quantum Information} (Cambridge: Cambridge University Press)

\bibitem{EPR} Einstein A, Podolsky B and Rosen N 1935 \textit{Phys. Rev.} {\bf 47} 777


\bibitem{vedral0} Vedral V 2008 \textit{Nature} {\bf 453} 1004. For a review, see Amico L, Fazio R, Osterloh A and Vedral V 2008 \textit{Rev. Mod. Phys.} \textbf{80} 517


\bibitem{vedral2} Wie\'{s}niak M, Vedral V and Brukner \v{C} 2005 \textit{New J. Phys.} \textbf{7} 258

\bibitem{vedral1} Wie\'{s}niak M, Vedral V and Brukner \v{C} 2008 Phys. Rev. B \textbf{78}, 064108

\bibitem{wootters1} O'Connor K M and Wootters W K 2001 Phys. Rev. A \textbf{63}, 052302

\bibitem{griffith} D. J. Griffiths 2006 \textit{Introduction to Quantum Mechanics} (Pearson Education Inc.)

\bibitem{bose01} S. Bose 2003 Phys. Rev. Lett. \textbf{91} 207901

\bibitem{bose2} Arnesen M C, Bose S and Vedral V 2001 Phys. Rev. Lett. \textbf{87}, 017901



\bibitem{indrani} Bose I and Tribedi A 2005 Phys. Rev. A \textbf{72} 022314


\bibitem{sarthour1} Souza A M, Soares-Pinto D O, Sarthour R S, Oliveira I S, Reis M S, Brand\~{a}o P and dos Santos A M 2009 Phys. Rev. B \textbf{79} 054408

\bibitem{rappoport} Rappoport T G, Ghivelder L, Fernandes J C, Guimaraes R B and Continentino M A 2007 Phys. Rev. B \textbf{75} 054422

\bibitem{sarthour2} Souza A M, Reis M S, Soares-Pinto D O, Oliveira I S and Sarthour R S 2008 Phys. Rev. B {\bf 77} 104402 

\bibitem{dipta} Das D, Singh H, Chakraborty T and Mitra C 2013 New J. Phys. \textbf{15} 013047

\bibitem{bruckner} Brukner \v{C}, Vedral V and Zeilinger A 2006 Phys. Rev. A \textbf{73}, 012110


\bibitem{bonner} Bonner J C, Friedberg S A, Kobayashi H, Meier D L and Blote H W J 1983 Phys. Rev. B \textbf{27}, 1

\bibitem{xu} Xu G, Broholm C, Reich D H and Adams M A 2000 Phys. Rev. Lett. \textbf{84} 4465

\bibitem{Myers} Myers B E, Berger L and Friedberg S A 1969 J. Appl. Phys. \textbf{40}, 1149



\bibitem{mitra1} Panigrahi P K and Mitra C 2009 J. Indian Institute of Science \textbf{89}, 3 

\bibitem{Diederix} Diederix K M, Bl\"{o}te H W J, Groen J P, Klaassen T O and Poulis N J 1979 Phys. Rev. B \textbf{19}, 1



\bibitem{berger} Berger L, Friedberg S A and Schriempf J T 1963 Phys. Rev. \textbf{132} 1057


\bibitem{bandyo} Wu L -A, Bandyopadhyay S, Sarandy M S and Lidar D A 2005 Phys. Rev. A \textbf{72}, 032309

\bibitem{vedral3} Brukner  \v{C} and Vedral V, arXiv quant-ph/0406040.

\bibitem{dowling} Dowling M R, Doherty A C and Bartlett S D 2004 Phys. Rev. A \textbf{70}, 062113

\bibitem{toth} Toth G 2005 Phys. Rev. A \textbf{71}, 010301(R)

\bibitem{wang} Wang X 2002 Phys. Rev. A \textbf{66}, 034302

\bibitem{subir} Sachdev S 2000 \textit{Quantum Phase Transition} (Cambridge: Cambridge University Press)

\bibitem{philipp} Gegenwart P, Si Q and Steglich F 2008 Nature Phys. \textbf{4}, 186

\bibitem{osborne} Osborne T J and Nielsen M A 2002 Phys. Rev. A \textbf{66}, 032110

\bibitem{osterloh} Osterloh A, Amico L, Falci G and Fazio R 2002 Nature \textbf{416}, 608

\bibitem{vidal} Vidal G, Latorre J I, Rico E and Kitaev A 2003 Phys. Rev. Lett. \textbf{90}, 227902

\bibitem{venuti} Venuti L C, Boschi C D E, Roncaglia M and Scaramucci A 2006 Phys. Rev. A \textbf{73} 010303(R)

\bibitem{song} Song H F, Rachel S and Hur K L 2010 Phys. Rev. B \textbf{82} 012405

\bibitem{grenier} Grenier B, Boucher J -P, Henry J -Y, Regnault L -P and Ziman T 2007 \textbf{310} 1269. For a review, Giamarchi T, R\"{u}egg C and Tchernyshyov O 2008 Nature \textbf{4} 198-204

\bibitem{wootters2} Wootters W K 1998 Phys. Rev. Lett. {\bf 80}, 2245


\bibitem{friedberg} Friedberg  S A and Raquet C A 1968 J. App. Phys. \textbf{39}, 2


\bibitem{esr} Gopal E S R 1996 \textit{Specific Heat At Low Temperatures} (Plenum Press, New York)


\bibitem{blote} Bl\:{o}te H W J, Cardy J L and Nightingale M P 1986 Phys. Rev. Lett. \textbf{56} 742
\bibitem{song} Song H F, Rachel S and Hur K L 2010 Phys. Rev. B \textbf{82} 012405




\end{thebibliography}
\end{document}